\documentclass[a4paper,fleqn,usenatbib]{mnras}

\usepackage[T1]{fontenc}
\usepackage{ae,aecompl}

\usepackage{graphicx}	% Including figure files
\usepackage{amsmath}	% Advanced maths commands
\usepackage{amssymb}	% Extra maths symbols
\usepackage{ulem}

\newcommand{\ver}{VER J2016+371}
\newcommand {\pizero} {$\pi^{0}$-decay}
\newcommand{\brems}{bremsstrahlung}
\newcommand{\fermi}{\textit{Fermi-LAT}}
\title[TeV gamma-ray source \ver]{Implications of the pulsar wind nebula scenario for a TeV gamma-ray source \ver}

\author[Lab Saha]{
Lab Saha\thanks{E-mail: labsaha@ncac.torun.pl}
\\
Nicoulas Copernicus Astronomical Center, Rabia\'{n}ska 8, 87-100, Toru\'{n}, Poland\\
}

\date{Accepted 2016 May 23. Received 2016 May 23; in original form 2016 March 22}

\pubyear{2016}

\begin{document}
\label{firstpage}
\pagerange{\pageref{firstpage}--\pageref{lastpage}}
\maketitle

\begin{abstract}
We present multiwavelength studies of a TeV gamma-ray source \ver\ suggested to be associated with a supernova remnant CTB 87  (G74.9+1.2) and based on X-ray and radio morphologies, CTB 87 is identified as an evolved pulsar wind nebula.  A source in the vicinity of \ver\ is  also detected at GeV energies by Fermi Gamma Ray Space Telescope  suggesting a likely counterpart at GeV energies. We find that a broken power-law (BPL)  distribution of electrons can explain the observed data at radio, X-ray and TeV energies, however, is not sufficient to explain the data at MeV--GeV energies. A Maxwellian distribution of electrons along with the BPL distribution of electrons in low magnetic fields can explain the observed multiwavelength data spanned from radio to TeV energies suggesting this as the most likely scenario for this source. We also find that although the hadronic model can explain the observed GeV--TeV data for the ambient matter density of $\sim 20~ \rm cm^{-3}$, no observational support for such high ambient density makes this hadronic scenario unlikely for this source.

\end{abstract}


%%%%%%%%%%%%%%%%%%%%%%%%%%%%%%%%%%%%%%%%%%%%%%%%%%

%==================

\section{Introduction}
Pulsar wind nebulae (PWNe) are considered to be potential Galactic sources of radiation from radio to  very high energy gamma-rays. The non-thermal emission from a pulsar wind nebula (PWN) is believed to result from synchrotron and inverse Compton radiation of the high energy particles (leptons) injected from a rotation-powered neutron star in the presence of magnetic field. Detection of such PWNe by the present generation of high energy gamma-ray telescopes (e.g. MAGIC, HESS, VERITAS)  at GeV--TeV energies have revealed them as likely candidates for  very high energy gamma-rays (see, e.g., \citealt{Gaensler_2006} for a review). Moreover, observed characteristics of many of the unidentified GeV sources by Fermi Gamma Ray Space Telescope (\fermi), also detected  at TeV energies, are  similar to those of well-known PWNe (see, e.g, \citealt{Kargaltsev_2013} for a review).

A TeV source \ver\ has been recently resolved at TeV energies by VERITAS telescope system \citep{Aliu_2014} as a point source. This source has been detected with a statistical significance of $\sim$ 5.8$\sigma$ with measured integral energy flux of $(8.2 \pm 3.4_{stat} \pm 2.9_{sys} ) \times 10^{-13} ~\rm erg~\rm cm^{-2}~\rm s^{-1}$ between 1 and 10 TeV \citep{Aliu_2014}. This is positionally coincident with a supernova remnant (SNR) CTB 87 (G74.9+1.2) which is a centrally brighten SNR with no evidence of an SNR shell \citep{Dickel_1975,Duin_1975,Wallace_1997}. Detailed analysis  of the X-ray data of  CTB 87 from \textit{Chandra} has discerned its morphology as an evolved PWN with a putative pulsar residing at southeast to the remnant center \citep{Matheson_2013}.  A \fermi\ GeV source 3FGL J2015.6+3709 is positionally close to \ver. Although it has been associated with a blazar B2013+370 behind the Galactic plane \citep{Kara_2012,Acero_2015}, GeV association with \ver\ can not be suppressed due to low angular resolution at GeV and TeV energies compared to radio and X-ray energies.

In a PWN scenario, observed emissions from radio to X-rays are normally explained by the synchrotron radiation process, whereas the observed fluxes at GeV--TeV energies are explained by  inverse Compton (IC) emission mechanism with synchrotron photons, dust photons, and cosmic microwave background (CMB) photons.  A well-known example of such systems is the Crab Nebula whose emission extends from radio to very high energy gamma-rays (see \citealt{Hester_2008} for a review).  The observed emission spectrum from the Crab Nebula is well explained by the synchrotron radiation and IC processes \citep{deJager_1992,Atoyan_1996,Hillas_1998,Bednarek_2003}. In general, it is believed that for the old PWNe the IC scattering with   dust photons and CMB (hereafter IC-CMB) photons are dominant process at high energies, however IC with synchrotron photons (hereafter SSC) arising from the same population of electrons becomes dominant  for young PWNe like the Crab Nebula. In an alternative scenario for electrons,  bremsstrahlung process can significantly contribute to photons at high energies depending upon the density of the ambient medium. In addition to electrons, high energy protons (primarily heavy nuclei) may be accelerated \citep{Atoyan_1996,Bednarek_1997,Bednarek_2003,Amato_2003}, which can produce  GeV -- TeV photons through decay of neutral pions ($\pi^0$s) produced in inelastic p-p collisions. Although the Crab Nebula is considered as a prototype PWN, the number of PWNe whose morphologies, energetics, spectral indices are quite different from that of the Crab Nebula, is continuously increasing, thus forming a different class of PWNe. Multiwavelength studies of these sources can  provide significant information about the injected particle spectrum, dominant emission processes and magnetic fields in the emission volume.

In this paper, we study \ver\ at MeV--GeV energies considering data from \fermi\ in the region around this. In addition, 
we study the implications of a scenario in which observed radio, X-ray, and GeV emission are considered to be associated with the TeV emission from \ver, and they arise from a PWN type source. We find that a simple power-law (PL) distribution of electrons is not sufficient to explain the observed spectrum at GeV--TeV energies.  The observed gamma-rays at TeV energies can be  partially explained by IC-CMB \footnote{In the IC-CMB contribution we have included both CMB photons and photons from interstellar radiation field (taken from \citealt{Mathis_1983}).} process. However, the observed fluxes at MeV--GeV energies cannot be fitted well with either by SCC or by IC-CMB  
processes indicating requirements for a different type of electron spectrum in the emission volume. We find that a  BPL electron distribution can explain the observed data at TeV energies well. However, MeV--GeV data remains unexplained with this BPL electron distribution. A Maxwellian population of electrons together with the BPL distribution of electrons can explain the  observed multiwavelength data well suggesting this as the most likely scenario for \ver. In addition to synchrotron and IC spectra, we also consider both \brems\ and \pizero\ processes to account for the observed data at GeV--TeV energies. Although \brems\ process can not explain the observed data at high energies, \pizero\ process can explain the GeV--TeV data well for an assumed ambient density of $\sim 20~ \rm cm^{-3}$ which is, however, unlikely for this PWN.
These scenarios and their implications are discussed more quantitatively in the following sections within the framework of a PWN scenario.

The rest of the paper is organized as follows; First, in Section \ref{sec:data_analysis} we discuss details of data analysis of \fermi.  In Section \ref{sec:model} we calculate the multiwavelength photon spectra to explain the observed multiwavelength data. We discuss the results and the implications resulting from the  multiwavelength studies 
in Section \ref{sec:results}. 
Finally, we summarize our results and conclude in Section \ref{sec:conclusion}.

\section{Data analysis and results} \label{sec:data_analysis}

\fermi\ data for \ver\ taken in the period between 2011-01-01 (MJD 55562) and 2015-08-01 (MJD 57030) are analysed in this study. All gamma-ray events  taken from a circular region of interest (ROI) with radius 15$^\circ$ centred at the position of RA(J2000) = 20$^h$ 16$^m$ 02$^s$ and Dec(J2000) = 37$^\circ$ 11\arcmin ~52\arcsec  are extracted. We select the events suggested for \fermi\ Pass 8 analysis for Galactic point sources using {\it gtselect} of Fermi Science Tools (FST; v10r0p5). In order to prevent event contamination at the edge of the field of view due to the bright gamma-rays from the Earth's limb, gamma-ray events with reconstructed zenith angles greater than 105$^\circ$ are rejected. We use standard binned likelihood analysis. For spectral analysis of the data, the gamma-ray events are binned in energy at 8 logarithmic steps between 100 MeV and 300 GeV. To correctly model the background we consider all the sources with in our ROI from the 3rd \fermi\ (3FGL) catalog. Since the point-spread function of LAT is large, we also consider sources from the region 10 degrees away from the ROI to account for emission at low energies \citep{Abdo-2009}. Considering this extended region of the sources, exposure map, which depends on orientation, orbit location, pointing direction and live time of the data accumulation, are produced. For spectral modelling, we use the newly introduced instrument response function {\it P8R2\_SOURCE\_V6}. The diffuse Galactic emission (gll\_iem\_v06.fits) and isotropic emission models (iso\_P8R2\_SOURCE\_V6\_v06.txt) are used  for the binned likelihood analysis using the \textit{gtlike} tool of FST. To determine the best set of spectral parameters of the fit, the parameters of the 3FGL sources within 3\degr around \ver\ are varied. We keep all the parameters of \fermi\ 3FGL sources fixed, which are more than 3\degr away from the center of the ROI.
In this work, we use  python based software {\it enrico} \citep{enrico-2013}  for \fermi\ analysis.

\begin{figure}
\includegraphics[scale=0.5]{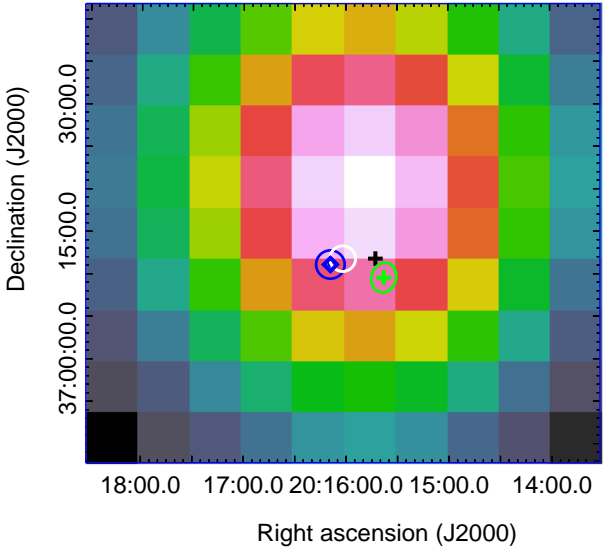}
\caption{Gamma-ray TS map of the source considered to be associated with \ver.  The best fit position of the 
source obtained with \textit{gtfindsrc} of Fermi tools is shown with a black cross. The green cross represents the 2nd Fermi- LAT catalog source with an error ellipse (green) with 95\% confidence. The blue diamond represents the best fit position of the putative pulsar with a blue circle for the diffuse nebula. The dashed white circle indicates the systematic uncertainty of $\sim$ 1\arcmin.5 \citep{Aliu_2014} in the best fit position of \ver. }\label{fig:tsmap}
\end{figure}

\begin{figure}
\centering
\includegraphics[scale=0.45]{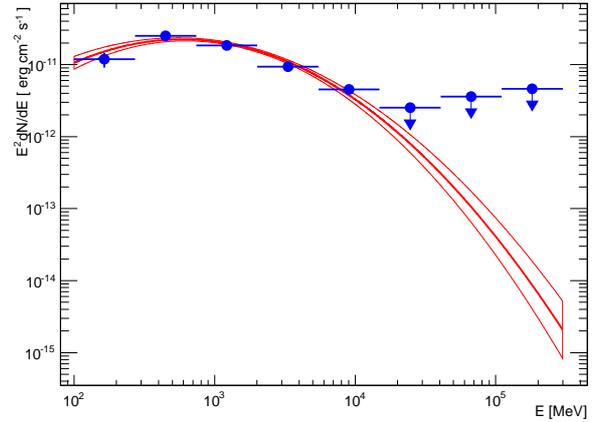}
\caption{\fermi\ spectrum of the source considered to be associated with \ver. The best fit curve along with  1$\sigma$ error bars are shown with solid lines. The parameters of the fit for LP model is given in the text. }\label{fig:fermi-sed}
\end{figure}

We have detected a source (positionally coincident with  VER J2016+371) with a statistical significance of $\sim$36 $\sigma$ using binned likelihood analysis. The best fit position within the ROI of \ver\ obtained using \textit{gtfindsrc} tool of FST is found to be longitude, l = 303.928 $\pm$ 0.01 and latitude, b = 37.1969 $\pm$ 0.01.  The model is then refitted using the best fit position to compute the TS map and differential energy spectrum. The TS map is shown in Figure \ref{fig:tsmap} with a cross (black)  which indicates the best fit position.  The  statistical positional uncertainty  of \ver\  estimated by VERITAS is shown with a dashed white circle.  The X-ray peak position is shown with a diamond (blue). The position of the source  3FGL J2015.6+3709 from 3FGL catalog is shown with a cross (green) with an error ellipse (green) of 95\% confidence.  The best fit position of the source obtained in this analysis is separated by 0.04\degr from the \fermi\ source 3FGL J2015.6+3709 and shifted towards the best fit positions at X-rays and TeV gamma-rays as clearly seen from Figure \ref{fig:tsmap}. 
The source spectral energy distribution (SED) of \ver\ at MeV--GeV energies is shown in Figure \ref{fig:fermi-sed} and  is best described by a log parabola (LP) function between 100 MeV and 300 GeV. The functional form of the LP is shown in Eqn. \ref{eqn:LP}.

\begin{eqnarray}
\rm {dF \over dE} = N_o (E/E_b)^{-(\Gamma_1 +\Gamma_2~ \rm ln(E/E_b))} 
\label{eqn:LP}
\end{eqnarray}
The best fit parameters for the LP model are $\rm N_o = (4.85 \pm
0.22) \times 10^{-12}~\rm MeV^{-1} ~\rm cm^{-2} \rm ~s^{-1}$ , $\Gamma_1$ = 2.45 $\pm$ 0.46, and $\Gamma_2$
= 0.23 $\pm$ 0.03 and $\rm E_b$ = 1522 MeV, where the given uncertainties are statistical. The total flux is found to be,   $\rm F(>100~ \rm MeV) = (1.01 \pm 0.01) \times
10^{-7} \rm photons ~\rm cm^{-2} ~\rm s^{-1}$.

\section{Multiwavelength modelling} \label{sec:model}
For multiwavelength modelling of \ver\ we use published radio fluxes at different radio frequencies (\citealt{Pineault_1990,Wendker_1991,Kothes_2006,Sun_2011}). X-ray data and GeV--TeV data are taken from \citet{Matheson_2013} and \citet{Aliu_2014}, respectively. For X-ray data, we consider the total X-ray flux $1.8 ~ \times 10^{-12} ~\rm erg ~cm^{-2}~ s^{-1}$ as an upper limit from the extended diffuse nebula of size $\sim 200\arcsec
 \times 300\arcsec$  \citep{Matheson_2013}.  
The observed fluxes at MeV -- GeV energies are taken from the results as given in Section \ref{sec:data_analysis}. The distance to the source is considered as 6.1 kpc \citep{Kothes_2003}.

We first consider a leptonic scenario to explain the observed data. For simplicity, a single population of electrons is considered. We assume a simple PL form of distribution of electrons ($\sim \gamma^{-\alpha}~\exp[-\gamma/\gamma_{max}$]) with a high energy cutoff at $\rm E_{max} = \gamma_{max} m_e c^2$.

In general, the electron spectrum may be  more complicated than the single power-law form. For the Crab Nebula, two different population of electrons are considered, namely, radio electrons and wind electrons. Radio electrons are less energetic electrons which reside in the nebular volume throughout its age, and they are mostly responsible for the observed radio fluxes. On the other hand, wind electrons are freshly accelerated electrons and they account for the observed fluxes at X-ray and GeV--TeV energies. In the case of the Crab Nebula, low energetic photons from radio synchrotron nebula are upscattered by the wind electrons giving rise to high energy photons at GeV -- TeV energies. Unlike Crab-like young PWNe, for relic PWNe, the high energy photons are produced by the up-scattering of CMB photons along with photons from stellar dust contribution since the density of radio photons is low in the emission volume.    
  
\begin{figure*}
\centering
\includegraphics[scale=0.8]{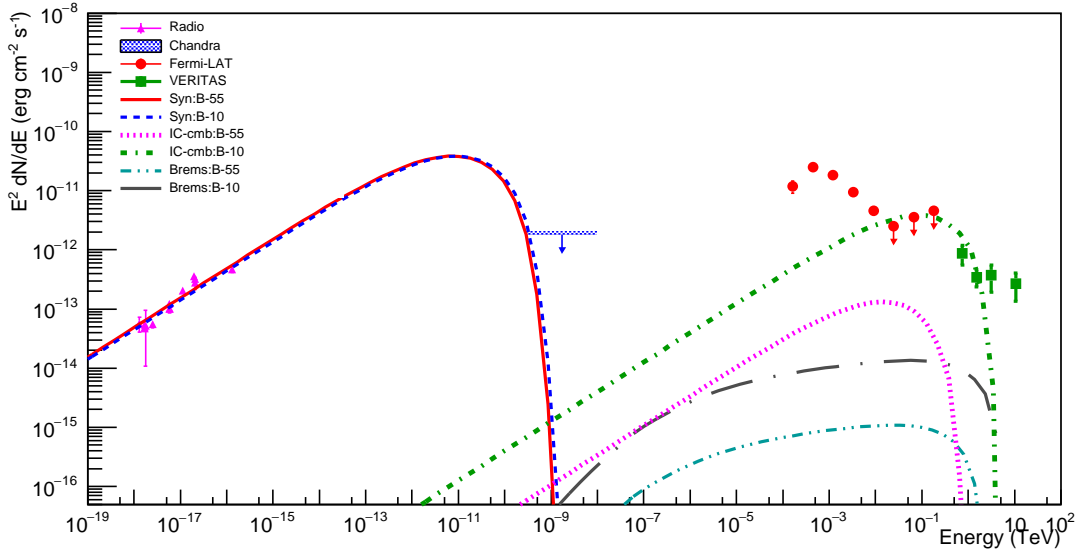}
\caption{The spectral energy distribution (SED) of \ver\ from radio to TeV energies in the PWN scenario for the magnetic fields of $55~\mu$G and $10~\mu$G. Radio data and X-ray data are explained by synchrotron spectra for both these magnetic field values as shown with solid and dashed lines (marked by the key ``B-55'' for B = 55 $\mu$G and ``B-10'' for B = 10 $\mu$G), respectively. The IC-CMB spectra for both these cases are shown with dotted and dot-dashed lines, respectively. The bremsstrahlung spectra for these magnetic field values are shown with a double-dot-dashed and long-dashed-dot  lines, respectively, for the ambient matter density of $0.2 ~\rm cm^{-3}$.}
\label{fig:spectral-fit}
\end{figure*}

In order to explain the observed spectrum for \ver, we first account for the observed radio fluxes for magnetic field (B) $\sim$ $55 ~\mu G$ as estimated by \citet{Matheson_2013}.  The calculated synchrotron spectrum is shown in Figure~\ref{fig:spectral-fit} and the parameters of the fit are shown in  
Table \ref{tab:fit_parameters_55muG}. The peak of the synchrotron spectrum depends on the maximum energy of the electrons, and very often it is restricted by the peak of the observed X-ray spectrum.  
In this case,  we  use the total X-ray flux as an upper limit in the energy band  0.3--10 keV. Hence, the maximum energy of the electrons cannot be defined well. However,  we choose the maximum energy of the electrons in such a way that the calculated fluxes at X-ray energies do not overestimate the observed fluxes.
In order to explain the GeV--TeV data, we calculate spectrum resulting from the IC-CMB mechanism, and we see that   this electron population is unable to explain the observed GeV--TeV data through  IC-CMB spectra as evident from Figure \ref{fig:spectral-fit}. For the magnetic field of 55 $\mu$G, the calculated fluxes are much less than the observed fluxes at GeV--TeV energies. It is also evident from Figure \ref{fig:spectral-fit} that the shape of the IC spectrum is quite different from the observed spectrum.  Hence, a simple power-law distribution is not sufficient to explain the observed fluxes at GeV--TeV energies.

Since, for evolved PWNe, the magnetic field strength in the nebular volume is normally considered to be much less ($\sim 5 --10 ~\mu G $) than that for young PWNe ($>100~ \mu$G), we also calculate synchrotron and IC  spectra for the magnetic field of 10 $\mu$G. The best fit spectrum, for this case, is also shown in Figure \ref{fig:spectral-fit} and parameters of this model is shown in Table \ref{tab:fit_parameters_55muG}. Although IC contribution is unable to  account for the GeV -- TeV data (see Figure  \ref{fig:spectral-fit}), it is significantly increased, which  suggests that the lower value of the magnetic fields in the  emission volume is preferred for this source and consistent with other evolved PWNe (\citealt{Slane_2010} and references therein).

Normally, in the leptonic scenario, non-thermal bremsstrahlung process is introduced to explain the data at high energies when IC process fails to do so. Since IC spectrum cannot explain the observed data for  the PL electron population, as shown above, we invoke bremsstrahlung process.  The density of the ambient medium, however, is required to calculate  the contribution from bremsstrahlung process. \citealt{Matheson_2013} estimated  the density of the medium to be $< 0.2 ~cm^{-3}$ based on the observed  absence of an SNR shell.  For this estimated density of ambient medium, bremsstrahlung process cannot explain the observed fluxes at high energies for both the scenarios with the magnetic field values 55 $\mu$G and 10 $\mu$G as shown in Figs.~\ref{fig:spectral-fit}. Since bremsstrahlung spectrum linearly depends on the density of ambient medium, a higher density (20 -- 100 $\rm cm^{-3}$)  can significantly increase the contribution to the level of GeV--TeV fluxes. However,  the shape of the spectrum does not match well with the observed one as evident from Figures~\ref{fig:spectral-fit}.

Since a PL electron spectrum cannot explain the observed data at GeV--TeV energies, we  consider a BPL type  electron distribution as given by 
\begin{eqnarray}
{dn_e \over d\gamma}  \propto \begin{cases}
 \gamma^{-\beta} ~\mbox{for} ~\gamma <\gamma_{br}  \\
  \gamma^{-\lambda} \exp{\left(-{\gamma \over \gamma_c} \right )} ~ \mbox{for}~ \gamma_{br}\leq \gamma \leq \gamma_c.                   
  \end{cases}
\label{eqn:pi0}
\end{eqnarray}
 
\begin{figure*}
\centering
\includegraphics[scale=0.8]{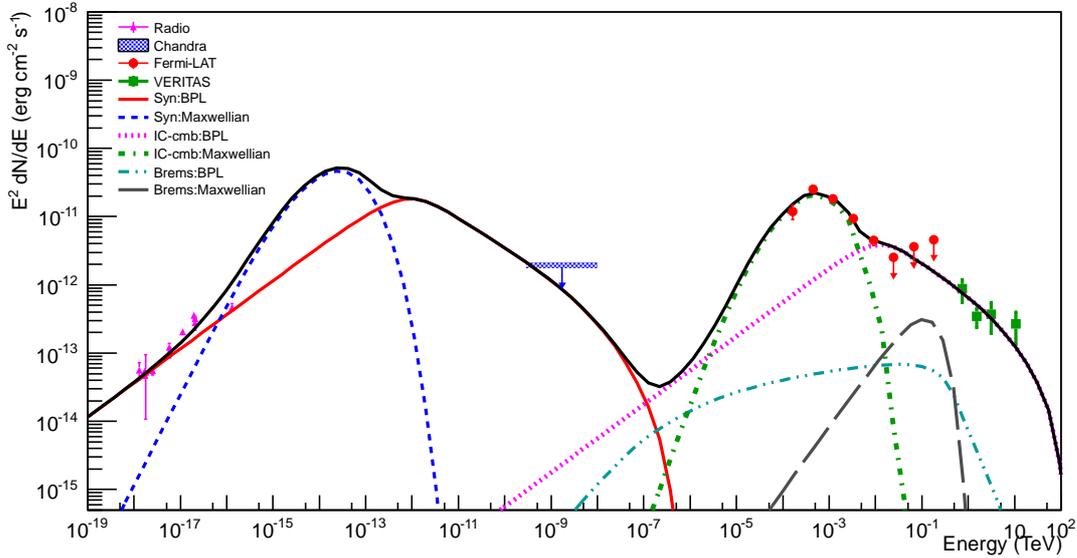}
\caption{Same as Figure \ref{fig:spectral-fit} but only for the different population of electrons. The synchrotron and IC spectra for a BPL distribution of electrons (marked by the key ``BPL'')   are shown with solid  and dotted  lines, respectively.   The dashed and dot-dashed lines  are synchrotron and IC spectra, respectively, for a Maxwellian distribution of electrons (marked by the key ``Maxwellian'') with mean $\gamma = 1.5 \times 10^5$.  The bremsstrahlung spectra for the BPL and Maxwellian electron distributions are shown with double-dot-dashed  and long dashed lines, respectively, for the ambient matter density of $0.2~ \rm cm^{-3}$. The thick solid  line  corresponds to combined fit to the data for these distributions of electrons.}\label{fig:spectral-fit_Max}
\end{figure*}

 We calculate the synchrotron and IC-CMB spectra for this type of electron distribution.
The model parameters are shown in  Table \ref{tab:fit_parameters_bpl} and corresponding SED is shown in Figure \ref{fig:spectral-fit_Max}. It is clearly illustrated in Figure \ref{fig:spectral-fit_Max} that the IC-CMB spectrum can explain the observed gamma-rays at TeV energies (the magnetic field is adjusted to $\sim~7~\mu$G), however, it underpredicts  fluxes at MeV -- GeV energies. Similar characteristics in the SEDs are observed in  two evolved PWNe Vela X \citep{Lamassa_2008,Grondin_2013} and HESS J1640-465 \citep{Slane_2010}, where  the simple power-law (or BPL) distribution of electrons fails to account for the observed fluxes, specifically, at MeV--GeV energies. 
To explain the observed \fermi\ data for HESS J1640 at MeV--GeV energies, a  Maxwellian distribution of electron population was chosen \citep{Slane_2010}. Such a particle spectrum was obtained in a particle-in-simulations study in the downstream of the wind termination shock \citep{Spitkovsky_2008}. Hence,  we consider a Maxwellian population of electrons ($\propto \gamma \exp[-\gamma/\delta\gamma] $) as an additional component to the BPL electron distribution. Figure \ref{fig:spectral-fit_Max} illustrates that a  Maxwellian distribution of electron with  $\delta\gamma = 1.5 \times 10^5$ can explain the observed flux well at \fermi\ energies,  where  BPL model fails. We also calculate bremsstrahlung spectra for both these electron distributions (BPL and Maxwellian), and we find that none of them can account for the observed fluxes for the ambient density of 0.2 $\rm cm^{-3}$.

As mentioned in the Introduction, although SSC is the dominant emission process for young PWNe, for evolved PWNe, it can contribute significantly when the emission region is considered very compact \citep{Saha_2015}. For this case, even if we consider that the emission is coming from the compact nebula of angular size $\sim$5\arcsec ($\simeq$ 0.15 pc at a distance 6.1 kpc), it is not sufficient for SSC to become a dominant process.

\begin{table}
\caption{Fit parameters  for a PL model for two different magnetic fields.}
\label{tab:fit_parameters_55muG}
\begin{tabular}{|c|c|c|}
\hline
parameters &     B = 55 $\mu$G  & B  = 10 $\mu$G \\
\hline 
spectral index  ($\alpha$)                 &   2.0             &   2.0     \\
Low energy cutoff ($\gamma_{min}$)         &   1.0             &   1.0     \\
High energy cutoff ($\gamma_{max}$)        &  3.5 $\times 10^6$  & 9.0 $\times 10^6$ \\
Total energy      ($10^{48}$ ergs)         &  1.13           & 14.1      \\
\hline
\end{tabular}
\end{table}

\begin{table}
\centering
\caption{Fit parameters  for a BPL model}
\label{tab:fit_parameters_bpl}
\begin{tabular}{|c|c|}
\hline
parameters &    values\\
\hline 
spectral index  ($\beta$)                 &   2.0  \\
spectral index (after break) ($\lambda$)                  &   3.8  \\
Low energy cutoff ($\gamma_{min}$) & 1.0 \\
High energy cutoff ($\gamma_c$)      &   3.3 $\times 10^8$ \\
Break position ($\gamma_{br}$) &   2.8 $\times 10^6$\\
Magnetic field ($B (\rm in ~\mu \rm G)$)  &   7    \\
Total energy    ($10^{49}$ ergs)            &  1.9 \\
\hline
\end{tabular}
\end{table}

\begin{figure}
\centering
\includegraphics[scale=0.45]{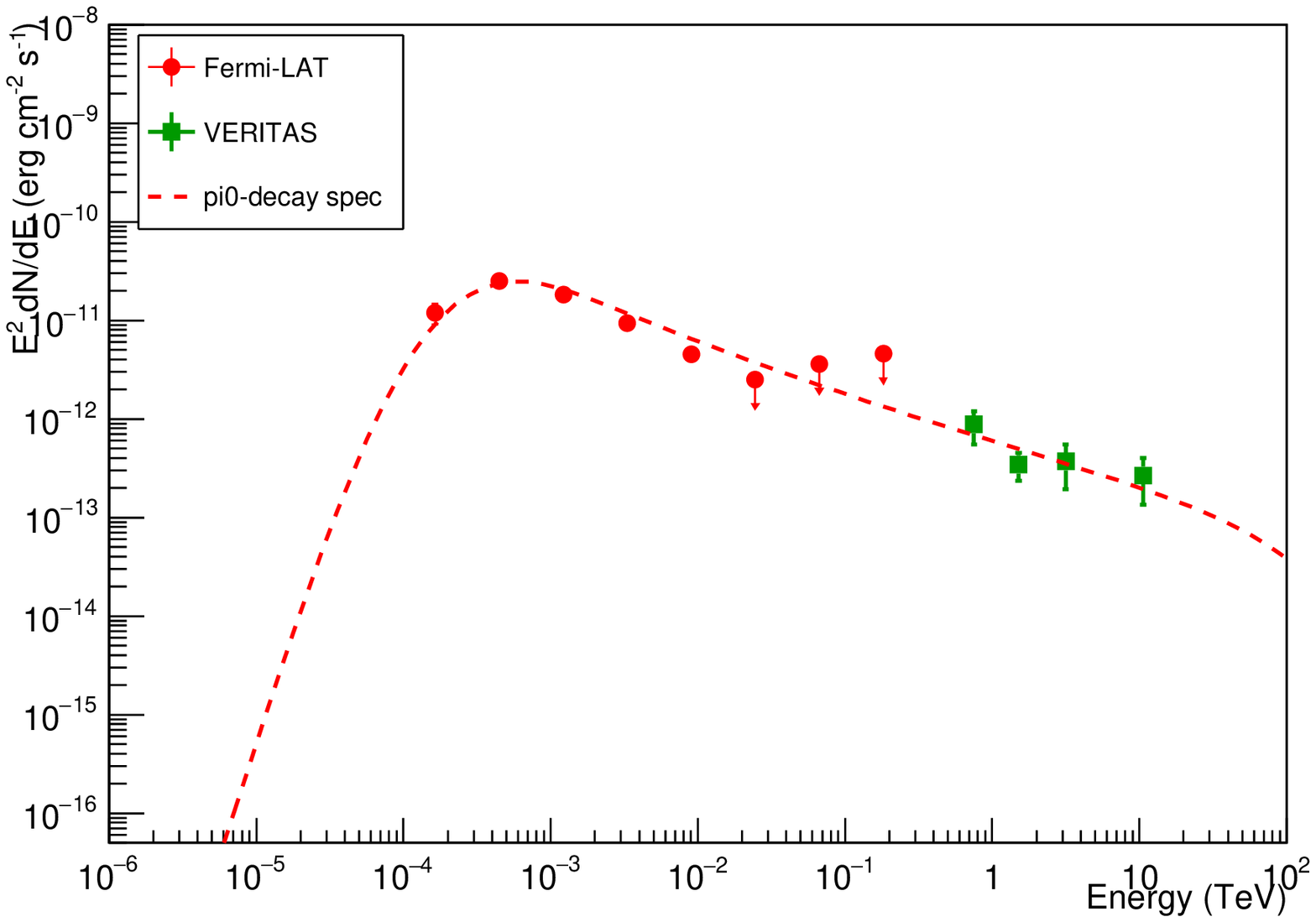}
\caption{Gamma-ray spectrum resulting from the decay of neutral pions. The best fit spectrum is calculated for the ambient proton density of 20 $\rm cm^{-3}$.}
\label{fig:pi0-spec}
\end{figure}
In addition to the leptonic scenario, we also introduce hadronic scenario as an addition component which mostly contributes to very high energies (MeV--TeV). We calculate gamma-ray spectrum resulting from the decay of neutral pions  following \citealt{Kelner_2006}. The gamma-ray spectrum for the relativistic protons with $ dN/dE\rm \propto E^{-2.55}$ with a spectral break at 100 TeV for  an ambient gas density of $n_H$ $\simeq 20$ cm$^{-3}$ is shown in Figure \ref{fig:pi0-spec}. The total energy  can be calculated  as $W_p =  3.26 \times 10^{50} \times (20.0/n_H)$  ergs. It is evident from the figure that the gamma-ray spectrum resulting from the decay of neutral pions can explain the observed GeV-TeV data very well. However, this ambient density is much higher than that the limit on the density of ambient medium of  $0.2 ~cm^{-3}$ as suggested by \citep{Matheson_2013}.  In general, such high densities are found in dense molecular clouds. There could be two different  scenarios in SNR for the interaction of molecular clouds. In the first one, one can assume that the clumpy molecular cloud could be present inside the SNR volume leading to p-p collisions. In the second scenario, it is considered that the relativistic protons already escaped the acceleration region and interact with molecular cloud outside the SNR volume. 
In this case, the total energy of protons requires to be unreasonably higher than that mentioned above ($\sim 10^{50} ~\rm ergs$) since the accelerated protons will lose a significant amount of energy while escaping from the SNR volume. Although a reasonable value of the total energy  can be considered for the protons escaping the SNR volume, the ambient matter density in the molecular cloud has to be much higher than that mentioned above ($\sim 20~ \rm cm^{-3}$). 
Moreover, the second scenario needs supports from observation where GeV--TeV emission should come from a region different from the peak of radio and X-ray positions. Such observational supports for none of the cases mentioned above for the sceond scenario are present.  In case of the first scenario, the presence of the dense molecular cloud (with ambient matter density of $\sim 20~ \rm cm^{-3}$) in the SNR volume can significantly contribute to the fluxes at GeV--TeV energies. However, such a dense medium will  be responsible for limb-brightened morphology for this PWN, which is not observed so far. Hence, both the scenarios within the context of hadronic model are unlikely. Nevertheless, for a less dense medium contribution from this scenario to total observed fluxes to some extent cannot be ruled out.

\section{Discussion}\label{sec:results}
The observed multiwavelength spectrum of \ver\ is studied considering both leptonic and hadronic scenarios which reveal some interesting characteristics of this source. We have seen that a simple PL distribution of electrons cannot explain the observed data at MeV--TeV energies. We have also found that although a BPL type of electron distribution can explain the data at TeV energies, the observed spectrum at MeV--GeV energies remains unexplained. 
However, a BPL type electron distribution and a Maxwellian population of electrons together can explain data well at MeV--TeV energies suggesting this as most likely scenario for this source.
In the leptonic scenario, the magnetic field of $\sim$ 10 $\mu$G is more preferable than the magnetic field estimated to be 55 $\mu$G by \citealt{Matheson_2013}. 
From the spectral parameters of the fit to the data in the extended nebular region, it can be seen that the magnetic energy density is much lower than the particle energy density, which implies that the  equipartition of magnetic energy and particle energy is not obeyed in this PWN. 
This  is consistent with other evolved PWN such as Vela X and HESS J160-465, where equipartition between magnetic energy and particle energy is not valid suggesting that magnetic energy has already converted into particle energy over its long life time.
Spatial distribution of magnetic field in the nebular volume is considered constant for this study. However, the magnetic field may depend on the distance from the central region of the nebula. A constant magnetic field  throughout the nebular volume, however, is a good approximation. Moreover,  in the context of a leptonic scenario of an evolved PWN,  the shape of the electron spectrum changes due to adiabatic and IC loss, which in turn changes the photon spectrum at GeV--TeV energies \citep{Zhang_2008,deJager_2009}. We see that the IC processes cannot account for the observed spectrum in the scenarios with a PL  electron distribution making the consideration of the losses insignificant. However, for the BPL model, since IC contribution becomes relatively significant, there will be some effects on the spectral shape at TeV energies, which can be adjusted with a different choice of parameters. For an evolved PWN the magnetic field is considered to be much less ($\sim 10 \mu$G) than that considered for young PWN like the Crab Nebula ($\sim 125 \mu$G), which yields higher gamma-ray flux relative to X-ray and radio fluxes. For \ver\ the observed radio and X-ray fluxes are less than the fluxes at very high energies making it a likely PWN with the magnetic field of  few microgauss.

In addition to synchrotron and IC contribution to the observed fluxes, we have considered bremsstrahlung process which can contribute to high energies. For the estimated matter density of $0.2~\rm cm^{-3}$, the bremsstrahlung contribution to high energy photons is not significant for any of the leptonic scenarios as mentioned above. Although bremsstrahlung spectrum can reach to the level of GeV--TeV fluxes for relatively higher values of ambient gas density, the shape of the spectrum is quite different from the observed spectrum for all the different electron distributions (PL, BPL, and Maxwellian) making it  a insignificant process for very high energy photons for this PWN.

In addition to the leptonic model, we have also considered hadronic model to explain the observed data at GeV--TeV energies.
Any compelling evidence for acceleration of protons in SNR can be obtained from the observed features in the spectrum at MeV --GeV energies, particularly when the spectrum falls steeply below $\sim$ 200 MeV \citep{Ackermann_2013}.
We have seen that the observed spectrum is very well fitted by the $\pi^0$-decay gamma-ray spectrum for an assumed matter density of $\sim ~20 \rm~ cm^{-3}$. The presence of molecular cloud in the close proximity of \ver\ can strongly support this scenario. \citealt{Kothes_2003} found evidence of an association of CTB 87 with a molecular cloud present towards the east of CTB 87. However,  the density may not be as high as $\sim ~20 \rm~ cm^{-3}$ since no limb-brightened morphology or any shell structure was seen for this source which can establish that the system evolves in a dense medium. Hence, hadronic scenario is less likely for this source. It is important to note that for a less dense medium hadronic process can contribute, although less significantly, to the total observed fluxes at MeV--TeV energies.

From the radio and X-ray observations, it is established that there is an offset of about 100$''$ from radio peak to X-ray peak \citep{Matheson_2013}. The electron distribution before the break in the BPL model is responsible for explaining the radio data and this is similar to less energetic radio electrons of the Crab Nebula. On the other hand, the electrons corresponding to the distribution after the break are more energetic and are responsible for explaining data at X-ray energies. This is also similar to the wind electrons as considered for the Crab Nebula. For the Crab Nebula, radio electrons and wind electrons are considered to be two different population from two different regions and they are responsible for observed fluxes from radio to gamma-rays\citep{Aharonian_1995, Meyer_2010}. In this case also, one can consider that two different population of electrons from  two different regions together form the BPL type electron distribution and this could be a possible reason for the shift between X-ray and radio peaks.

  We have considered the total X-ray flux from the diffuse nebula as an upper limit in the energy band of 0.3--10 keV for multiwavelength modelling of this source. However, the observed total X-ray spectrum from the PWN is fitted well with a power-law spectrum with spectral index $<2$ as reported by \citealt{Matheson_2013}.  Although the absence of any significant emission at low X-ray energies indicates that the observed X-ray flux is of non-thermal origin, contribution from emission of thermal origin cannot be ignored, for the limited sensitivity and field coverage of the X-ray observation \citep{Matheson_2013}. The presence of thermal-emission in the total observed fluxes  may significantly change the spectral shape of the X-rays. Hence, we consider the observed total flux as an upper limit. The observed X-ray spectrum with spectral index $< 2$ can be explained with the same population of electrons (PL or BPL ) which explains radio data, depending on the level of X-ray flux compared to radio flux and on the spectral shape of the radio spectrum. On the other hand, the X-ray spectrum with spectral index $>2$ can be easily explained with the falling edge of the synchrotron component of a PL or a BPL model.
We note that the observed spectral indices of the X-ray spectra of the nebula for both the evolved PWNe Vela X and HESS J1640-465 ($>$2) are quite harder than that for \ver\ ($<$2). As a result,  a single zone model is sufficient to explain the observed radio and X-ray data for Vela X and HESS J1640-465 \citep{Slane_2010,Lamassa_2008}.
It should also be noted that maximum or cutoff energy of the electrons can be normally obtained by the spectral steepening at X-ray energies. However, for this source, we do not see such shape in X-ray energies between  0.3 keV to 10 keV. Moreover, we have used total X-ray flux as an upper limit at these energies.  Thus, the maximum energy of electrons is not constrained. Nevertheless, IC mechanism can be used to restrict the maximum energy of electrons if the observed  fluxes are explained by this emission process (as in the case of BPL model). 
  
It is important to note that VERITAS source VER J2016+371 is positionally associated with 3FGL J2015.6+3709 which is considered to be associated with an FSRQ of unknown redshift \citep{Acero_2015}. Based on the variability index of the
Fermi-LAT source and its correlation with radio, \citealt{Kara_2012} associated the high energy gamma-ray emission with the nearby
blazar B2013+370, with unknown redshift. However, very high energy gamma-ray emission from this extragalactic object is not seen in the current VERITAS data \citep{Aliu_2014}, thus making this association unlikely. Moreover, the unknown blazar is separated by 6\arcmin.7 away from the centroid of VER J2016+371, which is much larger than $\sim$ 1.\arcmin5 uncertainty of the VERITAS measurement \citep{Aliu_2014}. In our present \fermi\ analysis, we have obtained best fit location of the source is about 0.04\degr away from the 3FGL J2015.6+3709 and it is towards the location of the PWN supporting the association of the \fermi\ source with CTB 87 as well as with \ver.

\section{Conclusion}\label{sec:conclusion}
We have seen that the observed spectrum from radio to TeV energies can be well explained by a leptonic scenario of a BPL  and a Maxwellian distribution of electrons when the observed TeV fluxes from \ver\ are associated with the positionally coincident counterparts at low energies. Specifically, the association of \fermi\ source with the TeV source indicates the presence of a Maxwellian distribution of electrons in the emission volume since the observed MeV--GeV fluxes are better explained by gamma-rays from IC process for this electron distribution. In addition, a hadronic scenario can also explain the observed GeV--TeV data for high density of the ambient medium. However, any strong observational evidences are not present to support this scenario as a significant scenario for this PWN. The dominant emission processes for \ver\ are obtained considering its association with the sources at radio and X-ray energies. 
Angular accuracy for the measurements at radio and X-ray energies are far better than that of measurements at gamma-ray energies. Hence, future gamma-ray instruments with far better angular resolution (e.g., CTA) can provide significant information required to understand the spectral and spatial structure of the source and validity of these associations.

\section*{acknowledgments}
Data obtained from the High Energy Astrophysics Science Archive Research Center (HEASARC), provided by NASA's Goddard Space Flight Center is used in this  work.  LS acknowledges the use of Fermi-LAT data and analysis tool from Fermi Science Support Center. LS would like to thank the anonymous referee for his/her comments and suggestions, which improved the manuscript. This work was  supported by the National Science Center (Poland), within the project DEC-2011/02/A/ST9/00256.

\bibliographystyle{mnras}
\bibliography{ver2016}

% Don't change these lines
\bsp	% typesetting comment
\label{lastpage}
\end{document}